\documentclass[a4paper,twocolumn,11pt, accepted=2024-04-10]{quantumarticle}
\pdfoutput=1
\usepackage[utf8]{inputenc}
\usepackage[english]{babel}
\usepackage[T1]{fontenc}
\usepackage{amsmath}
\usepackage{hyperref}
\usepackage[numbers]{natbib}
\usepackage{tikz}
\usepackage{lipsum}
\usepackage{mathtools}

\newcommand{\one}{1 \hspace{-1.0mm}  {\bf l}}

\newcommand{\of}[1]{\left(#1\right)}
\newcommand{\off}[1]{\left[#1\right]}
\newcommand{\offf}[1]{\left\{#1\right\}}
\newcommand{\mn}[1]{\langle #1\rangle}
\newcommand{\ket}[1]{\left| #1\right\rangle}
\newcommand{\bra}[1]{\left\langle #1\right|}
\newcommand{\braket}[2]{\left\langle #1\right|\left. #2\right\rangle}
\newcommand{\cd}[1]{c^\dagger_{#1}}
\newcommand{\cc}[1]{c_{#1}}
\newcommand{\der}[2]{\partial_{#1} #2}

\begin{document}
\title{Metrology and multipartite entanglement in measurement-induced phase transition}

\author{Giovanni Di Fresco }
\email[]{giovanni.difresco01@unipa.it}
\affiliation{Dipartimento di Fisica e Chimica ``Emilio Segr\`e", Group of Interdisciplinary Theoretical Physics, Universit\`a degli studi di Palermo, Viale delle Scienze, Ed. 18, I-90128
Palermo, Italy}
\author{Bernardo Spagnolo }
\email[]{bernardo.spagnolo@unipa.it}
\affiliation{Dipartimento di Fisica e Chimica ``Emilio Segr\`e", Group of Interdisciplinary Theoretical Physics, Universit\`a degli studi di Palermo, Viale delle Scienze, Ed. 18, I-90128
Palermo, Italy}
\author{Davide Valenti}
\email[]{davide.valenti@unipa.it}
\affiliation{Dipartimento di Fisica e Chimica ``Emilio Segr\`e", Group of Interdisciplinary Theoretical Physics, Universit\`a degli studi di Palermo, Viale delle Scienze, Ed. 18, I-90128
Palermo, Italy}
\author{Angelo Carollo}
\email[]{angelo.carollo@unipa.it}
\affiliation{Dipartimento di Fisica e Chimica ``Emilio Segr\`e", Group of Interdisciplinary Theoretical Physics, Universit\`a degli studi di Palermo, Viale delle Scienze, Ed. 18, I-90128
Palermo, Italy}
\maketitle
\begin{abstract}
Measurement-induced phase transition arises from the competition between a deterministic quantum evolution and a repeated measurement process. We explore the measurement-induced phase transition in the no-click limit of a one dimensional Ising chain through the quantum Fisher information (QFI) in two different metrological scenarios. In the first one, where we use an interferometric approach, we demonstrate through the scaling behavior of the QFI the transition of the multi-partite entanglement across the phases. In the second one, we consider the QFI with respect to the measurement strength. In this case, we show that the QFI signals the measurement induced criticality through its non-analytic behavior, in analogy to the fidelity susceptibility, in standard continuous phase transitions. Our results offer novel insights into the features of a quantum system undergoing measurement-induced phase transition, and indicate potential avenues for further exploration in the field of quantum physics.
\end{abstract}

\section{Introduction}
A many-body quantum system subject to a measurement process may undergo  abrupt changes, similar to quantum phase transitions (QPTs), as a function of the monitoring rate \cite{Rossini2020, Cao2019, Alberton2021,Biella2021,Turkeshi2021, Turkeshi2021, Turkeshi2023, Minato2022}. Unlike conventional QPT that are driven by external factors, these criticalities, called measurement-induced phase transitions (MIPTs), arise from the interplay between the deterministic unitary evolution and the measurement process. This phenomenon has been found to have important implications for understanding the properties of many-body quantum systems, such as the onset of quantum entanglement \cite{Gullans2020, Gullans2020-1,Chan2019, Skinner2019, Nahum2021, Choi2020, Sang2021} and the emergence of topological order~\cite{Lavasani2021}.\\
\indent In this work we propose to explore the concept of MIPT through the use of QFI. The reason to characterise this phenomenon through QFI is two-fold. On the one hand, the QFI is a measure of the sensitivity of a quantum system to small changes in a parameter. As such, the QFI has been widely exploited to signal and characterise the critical behavior of equilibrium~\cite{Mussardo2010, Zanardi2008,Invernizzi2008,Tsang2013,Ivanov2013,Bina2016,Frerot2018,Heugel2019,Garbe2020,Ivanov2020,Montenegro2021, Albarelli2022, Dicandia2021, DiFresco2022} as well as non-equilibrium QPTs~\cite{Banchi2014, Carollo2020}. It is thus natural to ask whether and how the QFI may respond to critical transitions when the latter are driven by measuring processes. On the other hand, one of the characterising features of MIPT is an observable qualitative change of the system's entanglement properties, which has been observed through the scaling behavior of the entanglement entropy \cite{Rossini2020, Cao2019, Alberton2021, Biella2021, Turkeshi2021, Turkeshi2023}. The QFI is a witness of multi-partite entanglement~\cite{Hyllus2012,Toth2012,Strober2014,Hauke2015}, as well known from quantum metrology~\cite{Helstrom1976, Szczykulska2016,Albarelli2020,Ballester2004, Vaneph2013,Genoni2013,Yuan2015a,Berry2015, Gessner2018,Rubio2019,Carollo2019,Albarelli2019, Sidhu2019,Tsang2020,Demkowicz-Dobrzanski2020}. The QFI thus provides a complementary characterisation of the quantum correlations accross MIPT, beyond that provided by the entanglement entropy. Indeed, on the one hand the QFI reveals the multi-partite nature of entanglement, on the other it distinguishes metrologically useful quantum correlations from other forms of entanglements. Moreover, QFI provides a means to characterize the entanglement of systems which are not necessarily in a pure state, a technical limitation which affects other measures, such as the entanglement entropy.\\
\indent Motivated by the above arguments, we examine the QFI in two complemetary metrological scenarios. In both of them we consider as a prototypical example the MIPT of a one-dimensional Ising chain in a transverse magnetic field and subject to continuous observations. We particularly focus our studies on the no-click limit. In the first scenario, we investigate the QFI associated to the system's response to a single-parameter unitary transformation. This metrological scheme represents the standard scenario for utilizing the scaling of QFI as a multipartite entanglement witness. In the second scenario, we study the time-dependent behavior of the QFI after the system undergoes a quench in the monitoring rate. In this scenario, our analysis uncovers a general exponential scaling of the QFI, but most importantly, the QFI displays a non-analytic singularity at the critical value of the measurement strength. The latter is a behavior similar to what often observed in equilibrium and non-equilibrium quantum phase transitions~\cite{Mussardo2010, Zanardi2008,Frerot2018, Banchi2014, Carollo2020}. Our findings provide new insights into both the behavior of quantum systems subjected to MIPT and the role of the QFI in characterizing the associated quantum criticality.

\section{Model}

We consider a one-dimensional Ising chain with a transverse magnetic field \cite{Lieb1961, Barouch1971, Isingforbeg}
\begin{equation}
H = -\sum_{i=1}^M\left[\sigma_i^x\sigma_{i+1}^x + h\sigma_i^z\right] \label{1}\,,
\end{equation} 
where $\sigma_i^{\alpha}$ ($\alpha=x,y,z$) are the usual Pauli operators on the $i$-th site. On top of the dynamics induced by the Hamiltonian $H$, each spin is subject to a measuring processes $n_i = \frac{1 + \sigma^z_i}{2}$, with measuring strength $\gamma$. The competition of the Hamiltonian evolution and the measurement process results in a dynamics that can be described by the following stochastic quantum jump differential equation \cite{Turkeshi2021}
\begin{equation}
\begin{split}
&d\ket{\psi} = -iHdt\ket{\psi_t} -\frac{\gamma}{2}dt\sum_{i}\of{n_i - \mn{n_i}_t }\ket{\psi_t} +\\
& \sum_i dN_t^i \of{\frac{n_i}{\sqrt{\mn{n_i}_t}  } -1} \ket{\psi_t},
\end{split}\label{2}
\end{equation}
where $dN_t^i$ are $M$ independent Poisson processes ($(dN_t^i)^2=dN_t^i$, $dN_t^i=0,1$) with ensemble averages $\overline{dN_t^i} = \gamma\mn{n_i}_tdt$. At each time step, the outcome $\{dN_t^i\}_{i=1}^M$ of the measuring process define one possible realisation of the stochastic evolution, a so called quantum trajectory. In this paper, we are interested in postselecting only the trajectory in which no jump occurs (no-click limit), i.e. such that $dN_t^i=0$ ($i=1\dots M$) for all time steps. This postselected trajectory evolves according to a deterministic dynamics generated by the following non-Hermitian Hamiltonian
\begin{equation}
H_{\mathrm{eff}} = H - \frac{i\gamma}{2}\sum_{i} n_i.\label{3}
\end{equation}
This model features a critical transition for $\gamma_c = 4\sqrt{1 - h^2}$ \cite{ Biella2021, Turkeshi2023, Lee2014}, which separates the region ($\gamma<\gamma_c$) where the entanglement scales logarithmically with the system size, from the region $\gamma>\gamma_c$ in which the entanglement is constant in the system size~\cite{Turkeshi2021, Turkeshi2023}.

\section{Entanglement properties}
The standard tool to detect MIPT is the entanglement entropy. This section aims at showcasing the efficacy and advantage of QFI as a tool to reveal such transitions and characterize the corresponding phases through their entanglement properties. One major advantage of QFI over entanglement entropy is that it can identify the presence of multi-partite entanglement in the state. The idea behind detecting entanglement through QFI relies on the scaling properties of the latter with the system size N. Indeed, it is well known that if the QFI scales as 
	\begin{align}\label{bound}
		F/N > m
	\end{align}
(with $m$ a divisor of $N$), then  there is at least a $(m+1)$-partite entanglement in the system~\cite{Hyllus2012,Toth2012,Strober2014,Hauke2015}.

We will apply this idea to the (pure) stable state of the system. To this end, we follow the procedure outlined in Ref.~\cite{Turkeshi2021} to derive the dynamics of our system. It is worth noting that the dynamics of the system is quadratic in the fermionic operators at all times, hence the stable state is a fermionic Gaussian state, which can be expressed as
\begin{equation}
\ket{\psi_t} = \frac{1}{\sqrt{\det\of{U_t}}}\exp\of{\frac{1}{2}\sum_{ij} Z_t^{ij}c_{i}^\dagger c_{j}^\dagger  }\ket{0},
\end{equation} 
with $Z_t = -\of{U_t^\dagger}^{-1}V_t^\dagger$, where
$U_t$ and $V_t$ are $N\times N$ matrices appearing in the following block representation of the $2N\times 2N$ matrix
\begin{equation}
\mathcal{U}_t = \of{\begin{matrix}
U_t & V_t^\dagger \\
V_t & U_t^\dagger
\end{matrix}}.
\end{equation}
In turn, $\mathcal{U}_t$ is the unitary matrix that transforms the original fermionic operators $\boldsymbol{c}=\of{c_1,\dots, c_N}^T$ into a set of (time-dependent) free fermion modes $\boldsymbol{\chi}_t=\of{\chi_1^t,\dots, \chi_N^t}^T$
\begin{equation}
\left(\begin{matrix} \boldsymbol{\chi_t} \\ \boldsymbol{\chi_t}^\dagger \end{matrix}\right) = \mathcal{U}_t\left(\begin{matrix} \boldsymbol{c} \\ \boldsymbol{c}^\dagger \end{matrix}\right)
\end{equation}
such that $\chi_i^t\ket{\psi_t} = 0$ $\forall i$. Hence $\ket{\psi_t}$, the state of the system at time t, is the vacuum of the time dependent fermion operators $\chi_i^t$. Thus, the dynamics of the system is entirely encoded in the operator $\mathcal{U}_t$, which can be derived through its Heisenberg equation of motion \cite{Turkeshi2021} 
\begin{equation}
d\mathcal{U}_t = \of{-iH_{\mathrm{eff}}dt}\mathcal{U}_t.\label{10}
\end{equation}
Meticulous preparation of the metrology scheme is required to utilize QFI as an entanglement witness \cite{Hauke2015}. We select open boundary conditions and set $h =0$ in Eq.~\eqref{1} as the presence of a transverse magnetic field does not affect the qualitative features of the transition results \cite{Turkeshi2021, Turkeshi2023}. The estimation protocol consists in a unitary transformation applied to the state of the system at time t,
\begin{equation}
\ket{\psi_t}_\varphi = e^{-i G \varphi}\ket{\psi_t}\label{11}
\end{equation}
 where $G$ is the Hermitian generator of the transformation and $\varphi$ the parameter to be estimated. We chose as initial state $\ket{\psi} = \ket{00...0}$\footnote{This state is chosen because it is a product state and is not entangled, making it optimal for studying the entanglement evolution of the system. The entanglement properties of the process are not affected by the initial state \cite{Turkeshi2021}. }. After a sufficiently long time, the state converges to a steady state. This is reflected in the QFI which stabilizes in the long time limit. We will focus on the QFI associated to the steady states.
One of the difficulties in using QFI to detect the entanglement is that the violation of the bound~\eqref{bound} provides a sufficient condition for multipartite entanglement. This means that the QFI in the above metrological scheme is not guaranteed to be sensitive to entanglement for an arbitrary value of $G$. It is indeed necessary to single out a suitable generator $G$ which is responsive to the critical change of the entanglement. Guided by the properties of the unperturbed Ising chain, a natural choice for $G$ is an operator proportional to a spin operators along a suitable direction $\hat{n}$, i.e. $S_n=\vec{S}\cdot\vec{n}$, where $\vec{S}=\of{S_x, S_y, S_z}$ with $S_\alpha=\sum_i\sigma_i^\alpha$ and $|\hat{n}|=1$. One can show that the effect of both $S_z$ and $S_y$ yields trivial results insensitive to entanglement, which leaves $S_x$ as the only reasonable choice. We find that $S_x$ as a generator\footnote{For clarity, we analyzed other generators, but we report only the results for $S_x$ as it is the most meaningful among those analyzed.} yields a metrological scheme able to resolve the different phases of the model, and characterize their entanglement properties. 
The QFI associated to Eq.~\eqref{11}, once we set $G = S_x$, is \cite{Ragy2016}
\begin{equation}
F= 4\Delta^2 S_x,\label{12}
\end{equation}
where $\Delta^2S_x$ is the variance of $S_x$, which can be evaluated with standard procedures~\cite{Barouch1971}. 
It is worth stressing that, although Eq.~\eqref{12} is typically studied in many-body systems near criticality, such as the fluctuation in the average magnetization, it acquires in the present context a completely different meaning, i.e. as a measure of entanglement. Incidentally,  the identification between the variance $\Delta S_x$ and QFI is only valid for a pure state system. These two quantities indeed depart from each other in a mixed state scenario~\cite{Liu_2020}.

\begin{figure}[h]
\includegraphics[scale= 0.55]{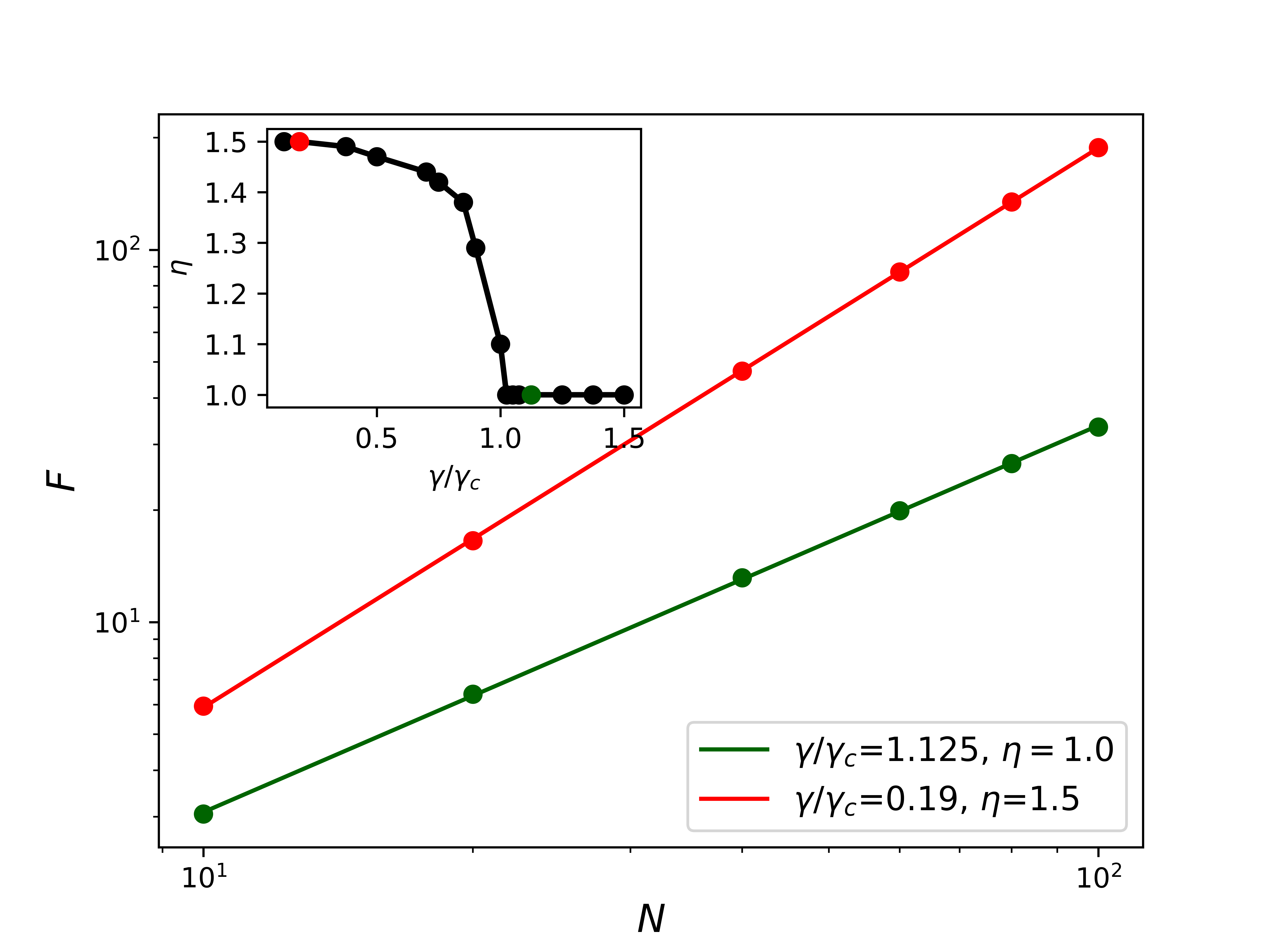} 
\caption{The log-log plot presented in this figure shows the scaling behavior of the QFI with $h=0$. The red line corresponds to the QFI behavior for $\gamma/\gamma_c = 0.19$, with an exponent of $\eta = 1.5$. In contrast, the green line shows the scaling behavior for $\gamma/\gamma_c = 1.12$, with an exponent of $\eta = 1$. The inset of the figure displays how the exponent $\eta$ of the scaling law changes as a function of $\gamma/\gamma_c$, highlighting a sharp transition at the critical point. The two colored points correspond to the lines reported in the main plot.}
\label{F1}
\end{figure}

The scaling of the QFI for the two distinct phases of the system is shown in Fig.~\ref{F1}. The QFI scaling for $\gamma <\gamma_c$ displays a super-extensive dependence, $F\propto N^\eta$ with $\eta=1.5$. In contrast, for $\gamma>\gamma_c$, the QFI follows a normal size-scaling behavior with $\eta=1$. The inset shows the sharp transition in the scaling law accross the phase transition. This shows that the scaling QFI is sensitive to this type of transition, displaying a qualitatively different behavior in the two phases. Moreover, the super-extensive behavior in the $\gamma<\gamma_c$ reveals the presence of a multipartite entanglement which is not observed in the $\gamma>\gamma_c$ phase.

This showcases the capability of the QFI to identify distinct phases across a MIPT, through the detection of multipartite entanglement.

Although the analysis carried out in this section focuses solely on the no-click limit,  this phenomenon is generally stable also when other trajectories are taken into account. Appendix C  describes how to incorporate jumps into the analysis and assesses the stability of this phenomenon for trajectories different from the no-click scenario, by showing that the transition is still present even when analysing different trajectories. 

\section{Non-Hermitian quantum quench}
This section aims to investigate the QFI of a system lying in the ground state of a Hermitian Ising Hamiltonian which is subject to a sudden "measurement quench". More specifically, the state is initially prepared in the ground state of the Hermitian Hamiltonian (Eq.~\eqref{1}), and from $t=0$ it is subject to a continuous measurement process, which results in an overall evolution described by Eq.~\eqref{2}. As in the previous section, we confine ourselves to the post-selected no-click trajectory, governed by the deterministic non-Hermitian Hamiltonian evolution $H_{\mathrm{eff}}$.
In this section we use periodic boundary conditions that allows us to exactly block diagonalize the Hamiltonian in $k-$space~\cite{Turkeshi2023, Lee2014}
\begin{equation}
H_{\rm eff} =\sum_k \left(\begin{matrix}c^\dagger_k & c_{-k} \end{matrix}\right)M_k\left(\begin{matrix}c_k \\ c_{-k}^\dagger \end{matrix}\right),\label{4}
\end{equation}
where $c_k$ are fermion annihilation operators obtained via Jordan-Wigner tranformation and Fourier transform of the spin operators $\sigma_i^{-}={(\sigma_i^x-i\sigma_i^y)/2}$ (see Appendix A), and
\begin{equation}
M_k = \left(\begin{matrix} \alpha_k & \beta_k\\ \beta_k & -\alpha_k \end{matrix}\right)\label{5}
\end{equation}
where $\alpha_k = -2\cos{k} -2h - \frac{i\gamma}{2}$, $\beta_k = 2\sin{k}$.
The spectrum of Eq.~\eqref{5}, $\epsilon_k = \pm \sqrt{\alpha_k^2 + \beta_k^2} = \pm E_k \pm i\Gamma_k$, has both real and imaginary part, which we denote as $E_k=\Re{(\epsilon_k)}$ and $\Gamma_k=\Im{(\epsilon_k)}$.  The steady state of the system is obtained by taking for each $k$ the eigenstate with negative imaginary part of $\epsilon_k$ \cite{Lee2014}. In the rest of the text, we will use the convention that $\epsilon_k$ has a negative imaginary part. The spectrum has a critical mode for $k=k_c\equiv\arccos\of{-h}$ whose eigenvalue, $\epsilon_{k_c}$, is real for $\gamma < \gamma_c$ and imaginary for $\gamma > \gamma_c$. 
Let us consider the state after a time evolution $t$ under the quenched measurement, which can be expressed as
\begin{equation}
	\ket{\psi_t} = N\of{t}e^{-iH_{\mathrm{eff}}t}\ket{\psi_0},\label{14}
\end{equation}
where $\ket{\psi_0}$ is the ground state of Eq.~\eqref{1} and $N\of{t}$ is a normalization factor. The state $\ket{\psi_t}$ will depend on the parameters $\gamma$ contained in the effective Hamiltonian $H_{\rm eff}$, and in particular on the strength of the measurement process $\gamma$. We will explore, through the QFI, the response of $\ket{\psi_t}$ to a small change of $\gamma$. Our goal is to demonstrate that the MIPT manifests itself as a non-analytical behavior of the QFI, a feature which is analogue to that observed across a second order QPT.
To compute the QFI for the state in Eq.~\eqref{14}, we extend the procedures of Ref.~\cite{Skotiniotis2015} to a non-Hermitian Hamiltonian. By using the standard definition of QFI for pure states, $
F = 4\of{ \braket{\partial_{\lambda} \psi_t }{ \partial_{\lambda} \psi_t } - \left\vert \braket{\partial_{\lambda} \psi_t}{ \psi_t} \right\vert ^2}$. We find that in a non-Hermitian scenario the QFI can be expressed in terms of the covariance of the operator $O_\lambda$ and $O_\lambda^\dagger$ calculated with respect to the state $\ket{\psi_t}$ (see Appendix B)
\begin{equation}
\begin{split}
&\tfrac{F}{4} =  \bra{\psi_t} O_\lambda^\dagger O_\lambda \ket{\psi_t} - \left\vert\bra{\psi_t} O_\lambda\ket{\psi_t}\right\vert^2
\end{split}\label{18},
\end{equation}
where the operator $O_\lambda$ is defined as
\begin{equation}
O_\lambda = \int_0^t ds \exp{\of{-iH_{\mathrm{eff}}s}}\der{\lambda}{\of{-iH_{\mathrm{eff}}}}\exp{\of{iH_{\mathrm{eff}}s}}.
\end{equation}
For simplicity, we assume periodic boundary conditions, which allows us to derive analytical expressions in $k-$space. In our setting, the parameter to be estimated is $\lambda=\gamma$. In this case, the operators $O_\lambda$ assumes a quadratic dependence on the fermionic operator $c_k$ and $c_k^\dagger$,  
\begin{equation}
O_\gamma = -\sum_k  \of{\begin{matrix}
\cd{k} & \cc{-k}\end{matrix}}
R_k\of{\begin{matrix}
\cc{k} \\
\cd{-k}
\end{matrix}} - \tfrac{Nt}{4},\label{n19}
\end{equation}
where $R_k$ is a $2\times 2$ time dependent matrix whose explicit, closed-form expression is given in Appendix B.
The state at time $t$ can be cast in a \emph{time-dependent} BCS-form~\cite{Turkeshi2023} as
\begin{equation}
\ket{\psi_t} =\prod_{k} \frac{u_k^t + v_k^t\cd{k}\cd{-k}  }{\sqrt{|u_k^t|^2 + |v_k^t|^2}}\ket{0},\label{N1}
\end{equation}
where $u_k^t$ and $v_k^t$ are time-dependent coefficients which can be expressed as
%
\begin{align}
	u_k^t &= u_k\cos\epsilon_k t\!-\! i\left(\tfrac{\beta_k}{\epsilon_k}v_k\right.
	\left.\!-\!\tfrac{\alpha_k}{\epsilon_k}u_k \right)\sin\epsilon_k t,\label{24-1}\\
	v_k^t &= v_k\cos\epsilon_k t \!-\! i\left(\tfrac{\beta_k}{\epsilon_k}u_k\right.
	\left.\!+\!\tfrac{\alpha_k}{\epsilon_k}v_k \right)\sin\epsilon_k t,\label{24-2}
\end{align}
and $u_k$ and $v_k$ are the standard coefficients of the ground state $\ket{\psi_0}$ of Eq.\eqref{1} (see Appendix A).
Using the above arguments, we derive the exact analytical expression for the QFI, which is explicitly reported in Appendix B. Fig.~\ref{F2} displays the behavior of the QFI as a function of time. It is observed that, following an initial transient period whose duration depends on $\gamma$, an exponential scaling regime emerges, with a scaling behavior approaching $2\gamma$. The exponential divergence observed is related to the non-Hermitianity of the model and it arises from the imaginary part of the spectrum. 
It is however not directly related to the critical features of the system.
\begin{figure}[h]
\includegraphics[scale=0.5]{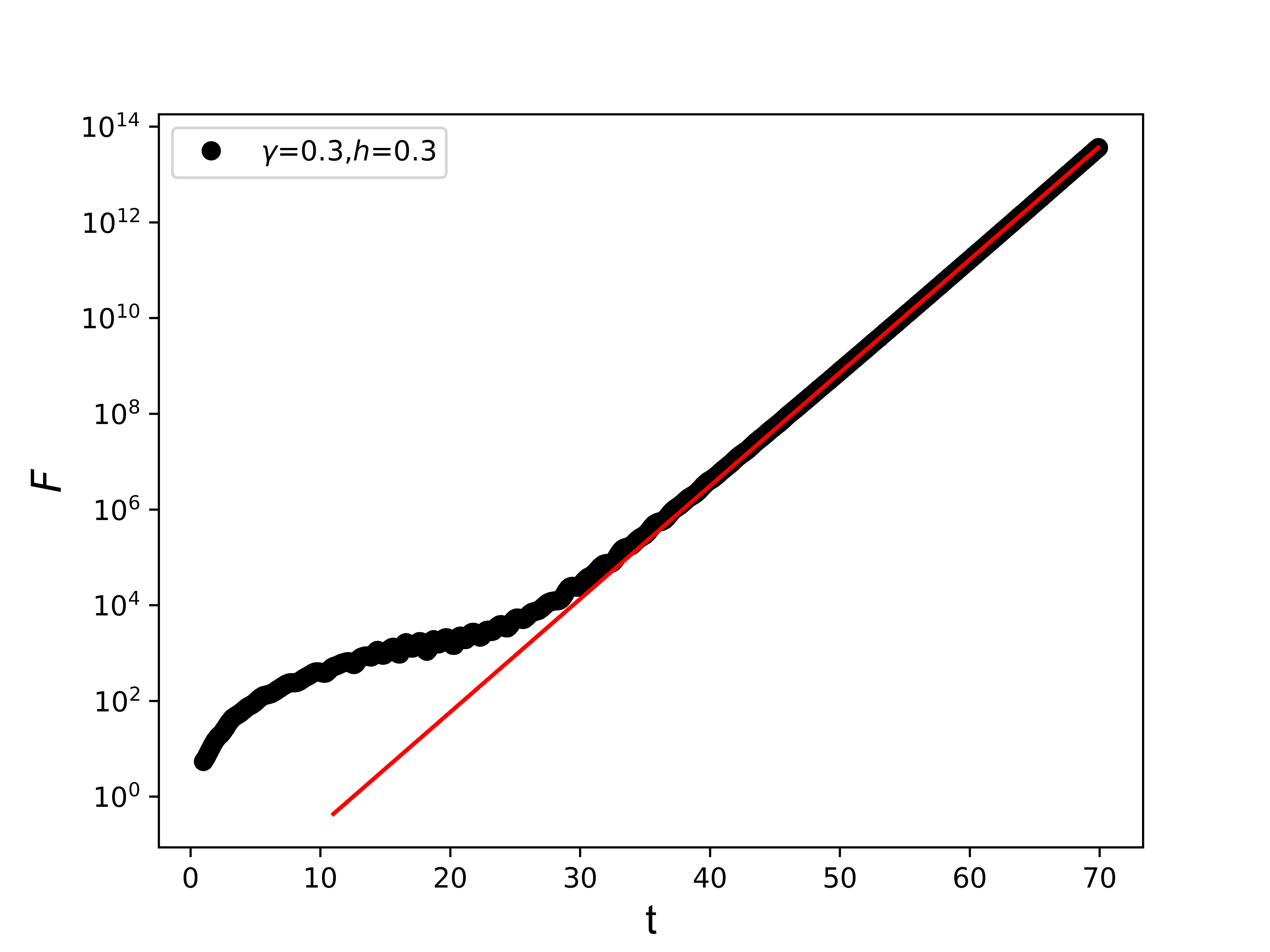} 
\caption{The figure presents the time evolution of the quantum Fisher information (QFI) for a set of parameters of the Hamiltonian, $h=0.3$ and $\gamma = 0.3$. We observe that an exponential scaling emerges, which depends on $\gamma$ and kicks in at a time that varies with the parameter values. The red line is an exponential fitting of the long-time behavior of the QFI.}
\label{F2}
\end{figure}

\begin{figure}[h]
\includegraphics[scale=0.5]{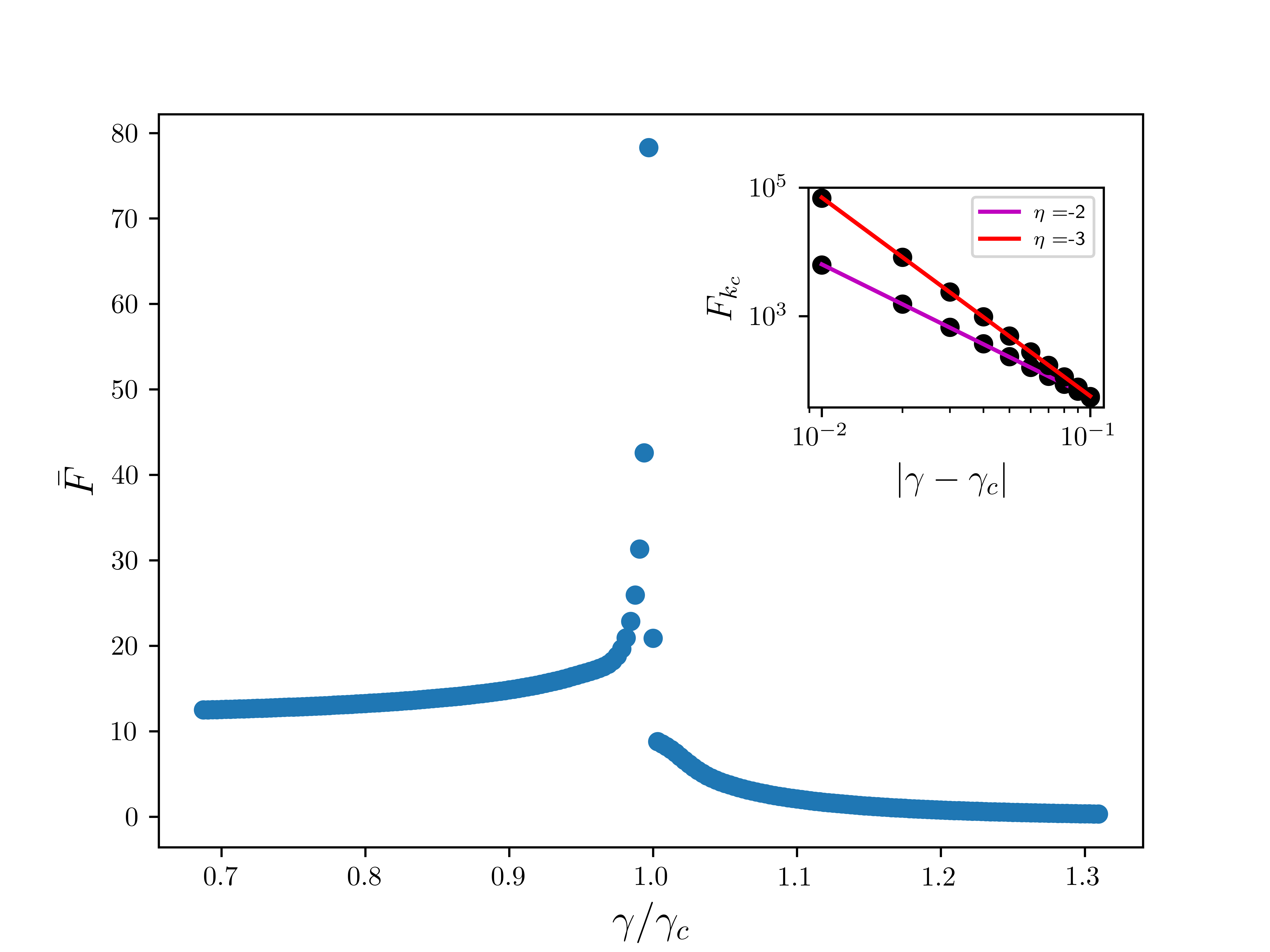} 
\caption{ The main figure shows the variation of the $\bar{F}$ for different values of $\gamma$, where $\gamma_c$ represents the critical value of the measurement rate (h = 0.6). In the inset, presented on a log-log plot, we depict the scaling behavior of the critical mode as reported in Eq.~\eqref{27} ($F_{k_c}\propto 1/|\gamma - \gamma_c|^{\eta}$), in magenta the scaling for $\gamma<\gamma_c$ and in red for $ \gamma>\gamma_c$. }
\label{F3}
\end{figure}

The effects of MIPT on QFI, such as possible non-analytic behaviors of QFI across a MIPT, are somewhat concealed by this exponential time dependence. To uncover the singular aspects of QFI, it is important to distinguish between the exponential time divergences and potential non-analytic divergences in $\gamma$. This distinction becomes evident when examining the long-time behavior of QFI, where one can decompose it into eigenmodes $k$, each primarily influenced by a specific time-dependent exponential factor
\footnote{Strictly speaking, there is a minor misuse of notation. In the case of $\gamma < \gamma_c$, the critical mode $k_c$ has a real eigenvalue $\epsilon_{k_c}$, resulting in a time dependence which is not exponential, but proportional to $t^2$. However, a careful analysis of this case does not impact the bounds in Eq.\eqref{N28}. This behavior is indeed taken into account explicitly in our examination of the critical mode's scaling.}
\begin{equation}
F_{t\rightarrow\infty}  = \sum_k F_k  e^{4\Gamma_k t}\label{N28}
\end{equation}
where, we recall, $\Gamma_k$'s are the imaginary part of the eigenvalues $\epsilon_k$, while $F_k$'s are time-independent coefficients, which  are expressed in detail in Appendix B. 
The $\Gamma_k$ functions are continuous with respect to $\gamma$, hence, any divergences of QFI due to the MIPT are to be found in the $F_k$ terms. This procedure provides a smooth mapping which leaves unaltered the analytic properties of the QFI. Indeed, one can explicitly relate any non-analytic divergent behavior of the QFI to that of an auxiliary function $\bar{F}=\sum_k F_k$ thanks to the following chain of inequalities
\begin{equation}
\bar{F} < F  < e^{\frac{\Gamma_{\rm{max}}}{4} t}\bar{F},\label{28}
\end{equation}
where $\Gamma_{\rm{max}}$ is the maximum value of the imaginary part of the spectrum, which is a  continuous function of $\gamma$.  This shows that $\bar{F}$ retains the same divergent behavior of $F$, but its study is considerably more amenable to numerical analysis, due to the absence of the exponential factors $e^{4\Gamma_k t}$.
In Fig.~\ref{F3} we plot the dependence of $\bar{F}$ on $\gamma$, which clearly displays a peak at the critical value $\gamma=\gamma_c$, with a distinctive asymmetric behavior as the criticality is approached from either above or below. 

We analytically derive the singular part of the QFI (see Appendix B), which scales as 
\begin{equation} F\sim F_{k_c}\propto
\left\{\begin{split}
&(\gamma - \gamma_c)^{-3}\hspace{0.5cm} \gamma>\gamma_c \\
&(\gamma_c - \gamma)^{-2} \hspace{0.5cm} \gamma<\gamma_c,
\end{split}\right.\label{27}
\end{equation}
 where $k_c = \arccos\of{-h}$ is a critical mode, i.e. the mode which is gapless at the MIPT. These features are in accordance with the numerical results shown in Fig. 3  and confirm them as a distinctive signature of the MIPT in the QFI behavior. The non-analytical behavior of QFI is strictly related to the closing of the gap at criticality. In particular, the different scaling laws as the critical point is approached from above or below $\gamma_c$ can be pin-pointed to the different behavior of the spectrum at the two sides of the criticality. The critical point is in fact an exceptional point \cite{Lee2014} of the model. This is associated to a gap of the model which is real for $\gamma < \gamma_c$ and imaginary for $\gamma > \gamma_c$. Heuristically, these qualitatively different behaviors of the spectrum reflect the distinct features of the two phases: (i) $\gamma < \gamma_c$, the imaginary part of the spectrum is gapless, and the correlation functions decay as a power law \cite{Turkeshi2021, Lee2014}; (ii) $\gamma > \gamma_c$, the dynamics is strongly influenced by measurement processes, leading to a dissipative, Zeno-like gap in which the correlation functions are expected to decay exponentially \cite{Turkeshi2021, Lee2014}. In both cases, the divergence of the QFI manages to capture the closure of the gap, and discriminates the distinctive features of the two phases through the asymmetry of the scaling laws. 
%
%
%
\section{Conclusion}
We have studied the impact of the MIPT on the QFI of a one-dimensional Ising chain subjected to a transverse magnetic field in the no-click regime. We have explored two distinct metrological scenarios. In the first scenario, where we can relate the scaling of the QFI to the entanglement properties of the system, our analysis reveals a change in the scaling behavior of the QFI across the transition point. We observe a super-extensive (sub-extensive) scaling law for values of the measuring rate $\gamma$ below (above) its critical value $\gamma_c=4\sqrt{1-h^2}$. This is revealed by the presence of multi-partite entanglement for $\gamma<\gamma_c$, not signalled for $\gamma>\gamma_c$. This behavior is consistent with the entanglement phase transition observed through the entanglement entropy \cite{Biella2021, Turkeshi2021, Turkeshi2023}. The QFI, however, provides a new insight into the multipartite nature of the entanglement associated with the two phases, demostrating its usefulness as a resource for quantum metrology.\\
\indent In the second one, we considered a sensing protocol in which the parameter to be estimated is encoded in the effective Hamiltonian. In this protocol, we show that the MIPT can be revealed through the divergent behavior of the QFI with respect to the measuring rate. This non-trivial finding parallels the analogous singular behavior of the classical and quantum Fisher information across equilibrium phase transitions, when the parameter estimated is that driving the transition \cite{Frerot2018}. Our findings provide new insights into the behavior of quantum systems subjected to MIPT, and open up the possibility of extending this approach to a general class of phase transitions in non-equilibrium physics, with competition between closed, open and measurement-induced dynamics. \\
\indent During the preparation of this manuscript the authors became aware of the related work~\cite{Paviglianiti2023}, which analyses multipartite entanglement across MIPT.

\textit{Acknowledgments---}
This work was supported by Italian Ministry of University and Research (MUR).
A.C. acknowledges financial support from Italian project PRIN $2022$-PNRR P$202253$RLY “Harnessing topological phases for quantum technologies.

The authors are grateful to Dr. Federico Roccati for useful discussions. 
\newpage
\onecolumn
\appendix

\section{Appendix A}
\begin{figure}
\centering
\includegraphics[scale= 0.6]{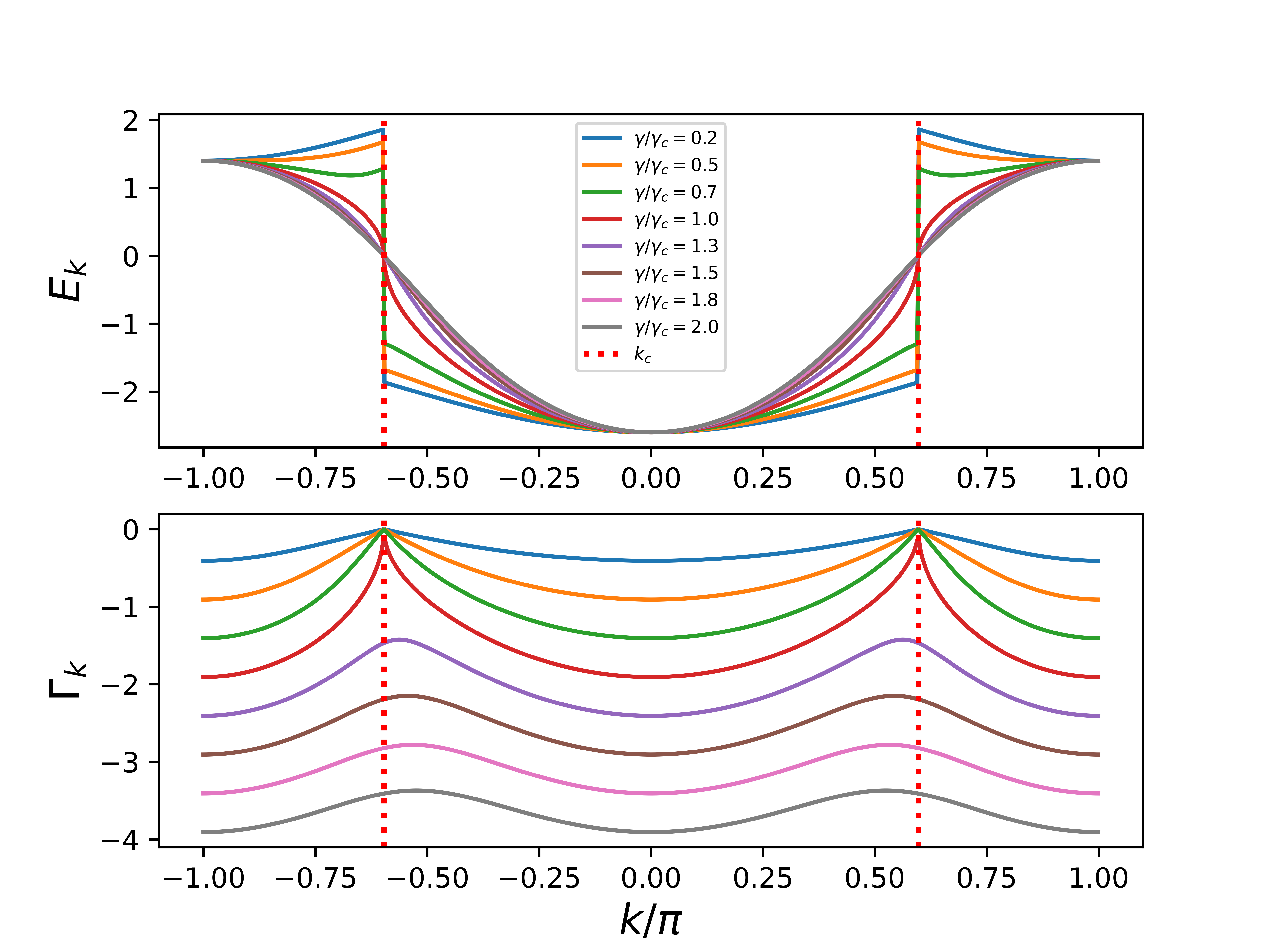} 
\caption{ The figure presents the spectrum of the Hamiltonian in Eq.~\eqref{bn1} for different values of the measurement rate $\gamma$. The upper panel displays the real part of the spectrum, gapped at $\pm k_c$ for $\gamma/\gamma_c<1$, while it becomes gapless for $\gamma/\gamma_c\geq1$. On the other hand, the lower panel shows the behavior of the imaginary part, which attains its maximum value $\Gamma_k = 0$ at $\pm k_c$ for $\gamma\leq\gamma_c$, whereas it remains non-zero for $\gamma >\gamma_c$.}
\label{F4}
\end{figure}

To ensure clarity and self-consistency to the work, in this Appendix we provide a set of information and details on the non-Hermitian Hamiltonian in Eq.~(3) of the main text. Firstly, we employ the Jordan-Wigner transformation
\begin{equation}
\left\{\begin{split}
&\sigma_i^x = \prod_{j<i} \of{2n_i -1}\of{c_i + c_i^\dagger}\\
&\sigma_i^z = 1 - 2c_i^\dagger c_i
\end{split}\right.
\end{equation}
to rewrite our Hamiltonian in fermionic language. Subsequently, we move to the momentum space through the Fourier transform $c_j=\frac{e^{i\pi/4}}{\sqrt{N}}\sum_k e^{ikj}c_k$. This results in the  expression of Eq.~(4) of the main text
\begin{equation}
H =\sum_k \left(\begin{matrix}c^\dagger_k & c_{-k} \end{matrix}\right)M_k\left(\begin{matrix}c_k \\ c_{-k}^\dagger \end{matrix}\right),\label{bn1}
\end{equation}
where
\begin{equation}
M_k = \left(\begin{matrix} \alpha_k & \beta_k\\ \beta_k & -\alpha_k \end{matrix}\right). \label{b1}
\end{equation}
In this work, every time we considered the expression of the Hamiltonian in the momentum space, we worked with periodic boundary conditions on the spin chain and an even number of fermions. This corresponded to use anti-periodic boundary conditions for the fermionic chain with 
\begin{equation}
k = \offf{ \frac{\of{2n -1}\pi}{N}\left\vert n= 1, ..., \frac{L}{2}\right. },
\end{equation}
where only the positive values were used due to the symmetries of the Hamiltonian. We also note that in Eq.~\eqref{bn1}, a constant term was neglected as it does not affect the system's dynamics. The spectrum of Eq.\eqref{b1}, denoted as $\epsilon_k = \pm \sqrt{\alpha_k^2 + \beta_k^2} = \pm E_k \pm i\Gamma_k$, possesses both real and imaginary parts. The real and imaginary parts are denoted as $E_k=\Re{(\epsilon_k)}$ and $\Gamma_k=\Im{(\epsilon_k)}$, respectively. It is important to note that the convention on the sign can be chosen independently for each $k$ in the system's spectrum. In this study, we have chosen the convention such that $\Gamma_k\leq 0$ always. Furthermore, it is worth noting that there exists a critical value of $k$, given by $k_c = \arccos\of{-h}$, where $\epsilon_k$ is real for $\gamma \leq \gamma_c$ and imaginary for $\gamma > \gamma_c$. This peculiar behavior of non-Hermitian systems is illustrated in Fig.\ref{F4}, where the behavior of the spectrum is reported for various values of $\gamma$.

The time-dependent state can be computed easily due to the initial state's well-defined parity and translation invariance. As a result, the dynamics can be decomposed into $N/2$ independent ones
\begin{equation}
\ket{\psi} = \prod_k \ket{\psi_k}.
\end{equation}
If we denote with $\ket{\tilde{\psi}} = \of{u^t_k + v^t_k\cd{k}\cd{-k}}\ket{0}$ the unnormalized $k$ state we can easily see that its dynamics is given by
\begin{equation}
i\frac{d}{dt}\of{\begin{matrix}
u^t_k\\
v^t_k
\end{matrix}} = M_k\of{\begin{matrix}
u^t_k\\
v^t_k
\end{matrix}},
\end{equation}
whose solution is given in Eq.~(19) of the main text, with initial condition given by
\begin{equation}
	\of{\begin{matrix}
		u_k\\
		v_k
	\end{matrix}} = \frac{1}{\sqrt{2\epsilon_k\of{\epsilon_k - z_k}}}
	\of{\begin{matrix}
		\epsilon_k - z_k \\
		y_k
	\end{matrix}},
\end{equation}
where $z_k = 2\of{h + \cos\of{k}}$ and $y_k = 2\sin\of{k}$.

\section{Appendix B}\label{appendix}
This Appendix presents the derivation of the QFI expression for the non-Hermitian quantum quench. As mentioned in the main text, the state at time $t$ can be expressed as a function of the initial state $\ket{\psi_0}$, using the general form
\begin{equation}
\ket{\psi\of{t}} = N\of{t} e^{-iH_{\mathrm{eff}}t}\ket{\psi_0}.
\end{equation}
To obtain the QFI, it is necessary to compute the derivative of the state with respect to a parameter of interest, which can be expressed as
\begin{equation}
\ket{\der{\lambda}{\psi}} = \of{\der{\lambda}{N\of{t}} e^{-iH_{\mathrm{eff}}t} + N\of{t}\der{\lambda}{e^{-iH_{\mathrm{eff}}t}}}\ket{\psi_0}.\label{A2}
\end{equation}
It is important to note that the derivative of the exponential term in the last term of Eq.~\eqref{A2} should be carefully considered, as $\partial_\lambda H$ and $H$ do not commute. The \textit{Sneddon's formula} \cite{Puri2001} allows us to express Eq.~\eqref{A2} as
\begin{equation}
\begin{split}
\ket{\der{\lambda}{\psi}}&=\of{ \tfrac{\der{\lambda}{N\of{t}}} {N\of{t}}  +\int_0^t ds \exp{\of{-iH_{\mathrm{eff}}s}}\der{\lambda}{\of{-iH_{\mathrm{eff}}}}\exp{\of{iH_{\mathrm{eff}}s}}} \ket{\psi}  \\
&=\of{ \tfrac{\der{\lambda}{N\of{t}}} {N\of{t}} + O_{\lambda}} \ket{\psi},
\end{split}\label{A3}
\end{equation}                                
and applying the general definition of the QFI for pure states, we can easily obtain
\begin{align}
\frac{F}{4} &= \braket{\partial_{\lambda} \psi }{ \partial_{\lambda} \psi } - \left\vert \braket{\partial_{\lambda} \psi}{ \psi} \right\vert ^2 \label{A4}\\
& = \bra{\psi}\left[\of{\tfrac{\der{\lambda}{N}  }{N}}^* + O_\lambda^\dagger  \right] \left[\tfrac{\der{\lambda}{N}}{N}  + O_\lambda\right]\ket{\psi} - \left[ \tfrac{\der{\lambda}{N}}{N}  + \bra{\psi} O_\lambda \ket{\psi} \right]\left[\of{\tfrac{\der{\lambda}{N}}{N}}^*  + \bra{\psi} O_\lambda^\dagger \ket{\psi}\right] \nonumber\\
& =  \bra{\psi} O_\lambda^\dagger O_\lambda \ket{\psi} - \bra{\psi} O_\lambda\ket{\psi}\bra{\psi} O_\lambda^\dagger\ket{\psi} \nonumber.
\end{align}
The closed expression for the operator $O_\gamma$ reported in the main text can be obtained by explicitly computing the derivative of the Hamiltonian with respect to $\gamma$, which yields the following result
\begin{align}
&O_\gamma =
-\int_0^t ds \hspace{0.1cm} e^{-iH_{\mathrm{eff}}s} \sum_{k} \of{\begin{matrix}
\cd{k} &\cc{-k}
\end{matrix}} \of{\begin{matrix}
1 & 0 \\
0 & -1
\end{matrix}} \of{\begin{matrix}
\cc{k}
\cd{-k}
\end{matrix} }e^{iH_{\mathrm{eff}}s}- \tfrac{Nt}{4}.\label{A5}
\end{align}
After evaluating the equation of motion for the operator $c_k$, we can rewrite Eq.~\eqref{A5} as 
\begin{equation}
O_\gamma = \sum_k  \of{\begin{matrix}
\cd{k} & \cc{-k}
\end{matrix}}\off{\int_0 ^t ds\hspace{0.1cm}e^{-iM_k s}   \of{\begin{matrix}
1 & 0 \\
0 & -1
\end{matrix}}e^{i M_k s}}
\of{\begin{matrix}
\cc{k} \\
\cd{-k}
\end{matrix}} -  \tfrac{Nt}{4}.
\end{equation} 
At this point, following a series of tedious but straightforward algebraic manipulations, we arrive at the conclusion that
\begin{equation}
O_\gamma = -\sum_k  \of{\begin{matrix}
\cd{k} & \cc{-k}
\end{matrix}} R_k\of{\begin{matrix}
\cc{k} \\
\cd{-k}
\end{matrix}} - \tfrac{Nt}{4},
\end{equation}
where
\begin{equation}
\begin{split}
R_k &=\of{ \begin{matrix}
A_k & B_k \\
C_k & -A_k
\end{matrix}}=\\ 
&=\of{ \begin{matrix}
 \frac{t\alpha_k^2}{\epsilon_k^2} + \frac{\sin\of{2\epsilon_k t}}{2\epsilon_k^3}\beta_k^2 & \frac{\alpha_k \beta_k t}{\epsilon_k^2} + \frac{2\beta_k \sin^2\of{\epsilon_k t}\epsilon_k i - \alpha_k \beta_k\sin\of{2\epsilon_kt}  }{2\epsilon_k^3} \\
 \frac{\alpha_k \beta_k t}{\epsilon_k^2} - \frac{2\beta_k \sin^2\of{\epsilon_k t}\epsilon_k i + \alpha_k \beta_k\sin\of{2\epsilon_kt}  }{2\epsilon_k^3} & -\frac{t\alpha_k^2}{\epsilon_k^2} - \frac{\sin\of{2\epsilon_k t}}{2\epsilon_k^3}\beta_k^2
\end{matrix}}\\ 
\end{split}\label{A7}
\end{equation}
Using the expression of Eq.~\eqref{N1} of the main text, we can evaluate the expectation value of Eq.\eqref{A4}, resulting in the following expression
\begin{equation}
\begin{split}
\frac{F}{4}&= \bra{\psi} O_\lambda^\dagger O_\lambda \ket{\psi} - \bra{\psi} O_\lambda\ket{\psi}\bra{\psi} O_\lambda^\dagger\ket{\psi} =  \\
&=\sum_k \left[\frac{\left[ |u^t_k|^2\of{|A_k|^2 + |C_k|^2} -2\Re\of{u^{t*}_k v^t_k \of{- A_kB^*_k + C_k A^*_k}}  + |v^t_k|^2\of{|A_k|^2 +|B_k|^2}\right]}{|u^t_k|^2 + |v^t_k|^2} +\right.\\
& -\left.\left\vert\frac{\of{ -|u_k^{t}|^2A_k - C_k u_k^{t*}v^t_k - B_ku^t_kv_k^{t*} + A_k |v^t_k|^2 }}{|u^t_k|^2 + |v^t_k|^2 }\right\vert^2\right].
\end{split}\label{A8}
\end{equation}

When examining the long-time behavior of the QFI, it becomes evident that we can factorize the exponential scaling. This can be seen from 
\begin{equation}
\left\{\begin{split}
&A_k = \frac{i\beta^2_k}{4\epsilon_k^3}e^{2i\epsilon_k t} = \tilde{A}_ke^{2i\epsilon_k t}, \\
& B_k = i\frac{-\beta_k\epsilon_k - \alpha_k\beta_k}{4\epsilon_k^3}e^{2i\epsilon_k t} =  \tilde{B}_k e^{2i\epsilon_k t},\\
& C_k = i\frac{\beta_k\epsilon_k + \alpha_k\beta_k}{4\epsilon_k^3}e^{2i\epsilon_k t} =  \tilde{B}_k e^{2i\epsilon_k t}.
\end{split}\right.\label{A9}
\end{equation}
These scaling laws are valid for all values of $k$ and $\gamma$, except for $k_c$ when $\gamma < \gamma_c$. To compute the long time behavior of the QFI, we can evaluate the expectation value by using the ground state of the effective Hamiltonian \cite{Lee2014}. We denote with $\tilde{u}_k, \tilde{v}_k$ its coefficients.
Then it follows that the long time behavior of the quantum Fisher information is
\begin{equation}
\begin{split}
&F = \sum_k \left[ \frac{\of{|\tilde{u}_k|^2\of{|\tilde{A}_k|^2 + |\tilde{C}_k|^2}  - 2\Re\of{\tilde{u}_k\tilde{v}_k^*\of{ \tilde{A}^*_k \tilde{C}_k  - \tilde{A}_k \tilde{B}^*_k   }} +|\tilde{v}_k|^2\of{|\tilde{A}_k|^2 + |\tilde{B}_k|^2}}e^{4\Gamma_k t} }{\of{|\tilde{u}_k|^2+ |\tilde{v}_k|^2 } } \right. +\\
& - \left. \frac{\left\vert\of{-\tilde{A}_k|\tilde{u}_k|^2 - \tilde{C}_k \tilde{u}_k^*\tilde{v}_k  -\tilde{B}_k\tilde{u}_k\tilde{v}^*_k +\tilde{A}_k|\tilde{v}|^2 }\right\vert^2 e^{4\Gamma_k t}}{\of{\of{|\tilde{u}_k|^2+ |\tilde{v}_k|^2 }}^2 } \right] =\\
&= \sum_k F_k  e^{4\Gamma_k t}.
\end{split}\label{A10}
\end{equation}
The critical mode's expression of the QFI can be determined analytically by evaluating  Eq.\eqref{A10} for $k_c = \arccos(-h)$. When $\gamma>\gamma_c$, the QFI can be computed using the expressions provided in Eq.\eqref{A9}. In contrast, when $\gamma<\gamma_c$, the linear term in $t$ becomes the leading term in the long time approximation of Eq.~\eqref{A7}. This phenomenon is responsible for the difference in the scaling laws of Eq.~(22) of the main text.

\section{Appendix C}
In this appendix, we verify whether our approach is able to discriminate the presence of a measurement-induced phase transition beyond the no-click limit. To this end, we consider the evolution along a generic trajectory, i.e. a trajectory where clicks can occur. The idea, is to simulate a random collection of quantum jump trajectories, to evaluate the quantum Fisher information associated to each of them, and average their values. Due to its non-linearity, the QFI averaged over each trajectory is generally different from the QFI of the averaged trajectory. This non-linearity is a necessary feature, which is indeed required for any witness of a measurement induced phase transition.
As it is standard, to simulate a single trajectory we divide the evolution in time steps of length $\delta t$. In each step, we evaluate  $p_i(t) = \gamma \delta t{\bra{\psi}} n_i \ket{\psi}_t$, which, for sufficiently small $\delta t$, can be interpret as the probability of observing a click during this time step on a random site $i$. Then, a Bernoulli random variable with success probability $p_i(t)$ is generated, whose outcome tells whether the jump has occurred. 
If this is the case, the state after the jump is updated to $\ket{\psi}_{t+\delta t} = \frac{n_i\ket{\psi}_t}{\sqrt{\mn{n_i}}_t}$. If no jump occurs the state is evolved according to the non-Hermitian Hamiltonian $\ket{\psi}_{t+\delta t} =(\one-iH_{eff}\delta t)\ket{\psi}_t$. This procedure is repeated for each step until the full evolution of a single trajectory is completed. Importantly, after each step the state remains Gaussian, whether or not it experiences a jump. One can therefore encode all the relevant information in the state's correlation matrix.

\begin{figure}
\centering
\includegraphics[scale= 0.6]{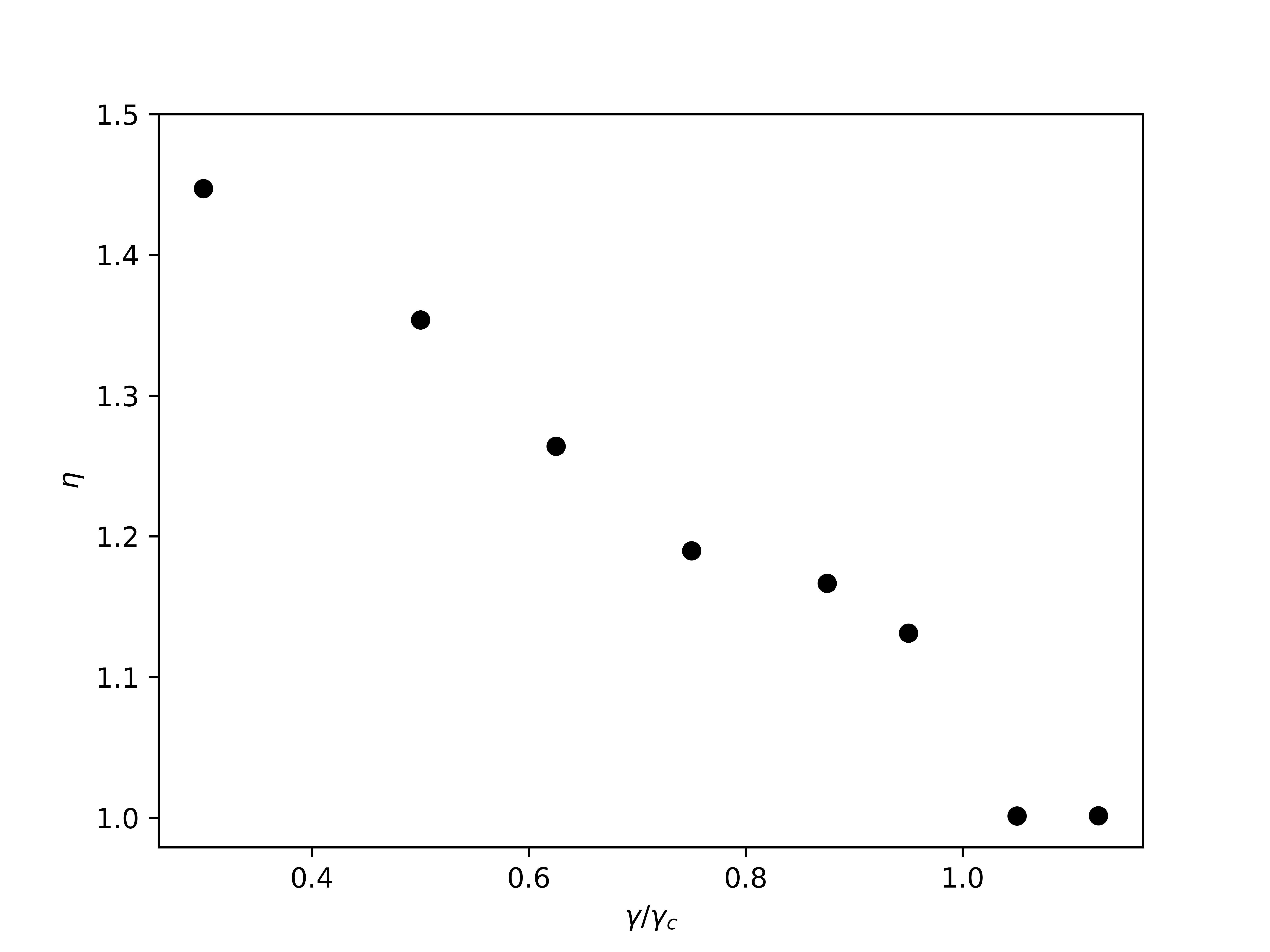} 
\caption{The figure illustrates how the exponent $\eta$ of the scaling law for the quantum Fisher information ($F\propto N^\eta$) changes as a function of $\gamma/\gamma_c$.}
\label{F5}
\end{figure}

In Fig.~\ref{F5}, we present the scaling exponent $\eta$ of the quantum Fisher information, $F_q \propto N^\eta$. To obtain this, we generated an ensable of 2000 trajectories for each $\gamma$ and for each different chain size $N$. Then, the QFI is averaged over all trajectories and the scaling law of the averaged QFI is determined in a  way similar to the no-click scenario. The plot in Fig.~\ref{F5} shows that abrupt changes in the scaling of the Fisher information are still present at a critical value $\gamma_c$, which value appears to be the same as that observed in the no-click scenario. Through this preliminary study, we show that the phenomenon of the measurement-induced phase transition remains present when jumps are allowed. However, due to the computationally more demanding scenario, we were able to demonstrate the scaling up to 20 spins.

As mentioned in the main text, the authors have recently become aware of related work \cite{Paviglianiti2023}, where the authors investigate scenarios involving clicks in the dynamics. A substantial agreement is observed with the result observed in \cite{Paviglianiti2023}.
\newpage
\bibliographystyle{quantum}
\bibliography{MIPT.bbl}

\end{document}